# Vertical Emission of Blue Light from a Symmetry Breaking Plasmonic Nanocavity-Emitter System Supporting Bound States in the Continuum

Yongqi Chen[1], Jiayi Liu[1], Jiang Hu[2], Yi Wang[2], Xiumei Yin[3], Yangzhe Guo[1], Nan Gao[1], Zhiguang Sun[1], Haonan Wei[1], Haoran Liu[1], Wenxin Wang[2,*], Bin Dong[3,*] and Yurui Fang[1, *]

[1.] *Center for Photophysics and Nanoscience, School of Physics, Dalian University of Technology, Dalian 116024, China.*
[2.] *Qingdao lnnovation and Development Center, Harbin Engineering University, Qingdao, 266500, China*
[3.] *Key Laboratory of New Energy and Rare Earth Resource Utilization of State Ethnic Affairs Commission, Key Laboratory of Photosensitive Materials & Devices of Liaoning Province, School of Physics and Materials Engineering, Dalian Nationalities University, Dalian 116600, China*

*Corresponding authors: wenxin.wang@hrbeu.edu.cn (W.W), dong@dlnu.edu.cn (B.D.), yrfang@dlut.edu.cn (Y.F.)*

**Abstract:**

The concept of photonic bound states in the continuum (BICs), introduced in structured metallic surface cavities, provides a crucial mechanism for designing plasmonic open-resonant cavities with high quality (high-Q) factors, making significant advances in plasmonic nanophotonics. However, the two major bottlenecks for plasmonic nanocavities: enhancing emission and big beam divergence for quantum emitters, due to the strong intrinsic Ohmic losses of metals. Here, we propose and realize a $\sigma_h$ symmetry-breaking plasmonic honeycomb nanocavities (PHC) that support quasi-BIC resonance modes with high-Q factors. Our anodic oxidation-engineered strategy breaks out-of-plane symmetry while preserving in-plane symmetry, enabling the PHC to exhibit collective plasmonic lattice resonances (PLR) couplings and achieve Q-factors exceeding $10^6$. Experimentally, we couple perovskite quantum dots (PQDs) to the PHC, demonstrating effective tuning of their emission properties and beam quality in the blue spectral region, achieving a 32-fold emission enhancement by suppress Ohmic loss and the life time of quantum emitters, simultaneously realize vertical emission in the 2.556 - 2.638 eV region, with a far-field hexagonal beam shape and a full width at half maximum of 12.6 degree under optimal coupling conditions. Furthermore, we demonstrate topological band inversion characterized by Zak phase transitions by continuously tuning the system parameters, confirming that the PHC supports topologically non-trivial q-BIC due to PLR coupling. The PHC presents itself as a promising next-generation, high-brightness nanoscale light source matrix, which can be directly scaled up to cover a wide wavelength range from UV to IR.

**Introduction**

Nanophotonics is a pivotal area of foundational research dedicated to manipulating and enhancing light-matter interactions. Benefiting from advanced nanofabrication techniques, a suite of profoundly impactful two-dimensional planar optical resonators has transitioned from theoretical models to experimental applications [1], including plasmonic cavities [2–5], [6–8], topological cavities [9–13], whispering gallery mode (WGM) resonators [14–16], and photonic crystal (PC) cavities [17–19]. These 2D platforms enable facile manipulation of light across multiple degrees of freedom and find broad applications in quantum information processing [20], biosensing [21,22], nonlinear optics [23,24], and laser technologies [12,25]. Particularly notable are plasmonic cavities, which spatially confine light fields to subwavelength regions and generate significant localized hotspots through their unique localized surface plasmon resonance (LSPR) at resonant wavelengths. Moreover, LSPR can further couple with Bloch states in periodic plasmonic structures, thereby leading to the formation of plasmonic lattice resonances (PLR) [26]. These quasi-particles serve as optical resonators beyond the diffraction limit, simultaneously hybridizing with semiconductor excitons to form nanocavity exciton polaritons, thereby enhancing light-matter interactions with a strong Purcell factor. Over the past few decades, plasmonic cavities have enabled breakthroughs in fields such as controllable photocurrents [27–29], low-dimensional materials [30], Raman spectroscopy [31] and optoacoustics [32,33]. However, in contrast to dielectric cavities, plasmonic cavities exhibit inherent large Ohmic losses and random scattering, resulting in resonant quality factors (Q-factor) typically ranging only from 10 to 100 [34,35], despite their small effective mode volumes $V_{\text{eff}}$. Moreover, plasmonic cavity systems either rely on highly ordered structures [36,37] or exist exclusively in non-uniform media [38], which constitute the primary barriers for practical application of plasmonic resonators.

Meanwhile, bound states in the continuum (BICs) [39], a non-Hermitian topological non-trivial state in an open system, has been investigated in plasmonic lattice cavities [40]. Different from localized surface plasmon resonances (LSPRs), which exhibit strong local resonance states capable of coupling to the far field, BICs exhibits a complete decoupling from radiative channels outside the system due to destructive interference mechanisms or symmetry mismatches. This leads to a theoretically infinite Q-factor and strong field limitations, even though this wave state lies in the continuum. Recently, researchers have witnessed the proliferation of hybridized gain media for integrating plasmonic cavity-emitter systems [8,41], exploiting the resonant strong coupling between photonic BICs and excitons to realize polarized exciton BICs both theoretically and experimentally. Indeed, the field enhancement achievable by the designed plasmonic lattice cavity at resonance is ultimately constrained by cavity losses and input coupling efficiency. The enhancement factor is described by the local field $E_{\text{loc}}$ and the input field $E_{\text{i}}$ [40,42]:

$$G = \frac{|E_{\text{loc}}|^2}{|E_{\text{i}}|^2} \approx \kappa_{\text{i}}^2 \frac{Q_{\text{tot}}^2}{V_{\text{eff}}} = \frac{Q_{\text{tot}}^2}{Q_{\text{r}} V_{\text{eff}}} = \frac{Q_{\text{r}}^2 Q_{\text{a}}^2}{Q_{\text{r}} (Q_{\text{r}} + Q_{\text{a}})^2 V_{\text{eff}}}, \qquad (1)$$

where $\kappa_{\text{i}}$ is the coupling coefficient with the external input field. Because it depends on the radiation channel, it can be represented as $\kappa_{\text{i}} = \sqrt{2\gamma_{\text{r}}}$, where $\gamma_{\text{r}} = \omega/(2Q_{\text{r}})$ denotes the radiation loss, and $\omega$ and $Q_r$ are the angular frequency and radiation quality factor, respectively. In lossy plasmonic structures, the total quality factor $Q_{\text{tot}}^{-1} = Q_r^{-1} + Q_a^{-1}$ is defined as a combination of radiative loss $Q_r^{-1}$ and non-radiative loss $Q_a^{-1}$, with $V_{\text{eff}}$ being the normalized effective mode volume. Analyzing from eq. (1), the cavity enhancement reaches its maximum under critical coupling conditions ($Q_r = Q_a$). For plasmonic structures, the mode volume $V_{\text{eff}}$ is typically small enough to provide substantial field enhancement. Recently, researchers achieved supercritical coupling by tailoring the edges where nanostructures meet the surrounding non-patterned substrate, enhancing exciton emission by up to eight orders of magnitude [42]. To date, abundant classical 2D planar cavity BICs designs aim to construct quasi-BICs (q-BICs) resonators with finite Q-factor by perturbing ideal BICs structures, e.g., creating in-plane symmetry-breaking geometries. Spatial disorder or oblique incidence light [43] can allow such q-BIC modes to couple with and be excited by the far-field. However, the achieved $Q_r$ of the actual structure is significantly diminished due to the high sensitivity of q-BIC resonance to in-plane perturbations. Moreover, the capability of the in-plane symmetry-breaking q-BIC cavity to control the direction of exciton emission is severely limited [44]. Therefore, the central question concerning the plasmonic cavity-emitter system is whether a viable design exists to achieve robust, high Q-factor plasmon cavities that amplify and directionally regulate radiation from quantum emitters.

    Here we present a blue-light vertically emitting plasmonic lattice cavity- emitter system supporting q-BIC. This system comprises a $\sigma_h$ symmetry-breaking honeycomb nanocavity array, which maintains in-plane $C_6$ symmetry while the tip nanoantenna supporting PLR. We reveal the transition from an ideal BICs to a q-BIC, moving from a triangular lattice with six-fold symmetry ($D_{6h}$) to a cone-shaped honeycomb lattice with three-fold rotational symmetry ($C_{3v}$), achieved by breaking the horizontal mirror symmetry ($\sigma_h$) of the geometrical structure. This physical mechanism opens the radiative channel, enabling the realization a high Q-factor dominated only by dissipation losses within the plasmonic system. We fabricated $\sigma_h$ symmetry-breaking PHC arrays using guided anodization and wet chemical etching. Using angle-resolved resonance scattering spectroscopy, we observed the vanishing of the PHC arrays spectrum at the BIC resonance. The calculated reflection spectrum verified this plasmonic BICs property, and the emission valley in the far field at the Γ point confirmed this $\sigma_h$ symmetry breaking q-BIC property. Furthermore, we experimentally demonstrated that the hybrid perovskite quantum dots-plasmonic honeycomb nanocavity system (Figure 1a) retains plasmonic BICs properties. The $\sigma_h$ symmetry-breaking honeycomb nanocavity effectively enhances the emission angle of the perovskite quantum dot luminescence. When a monolayer of perovskite quantum dots is deposited onto the plasmonic honeycomb nano-antenna, numerical simulations show that quantum dots located at different positions within the 2.5D lattice contribute differently to the overall emission enhancement, with those at the antenna tips contributing 16 times more than those at the base. Notably, we observe a Zak phase transition in the hybrid band under continuous variation of the system parameters, demonstrating the topological nature of this $\sigma_h$ symmetry-breaking q-BIC.

## Results and discussion

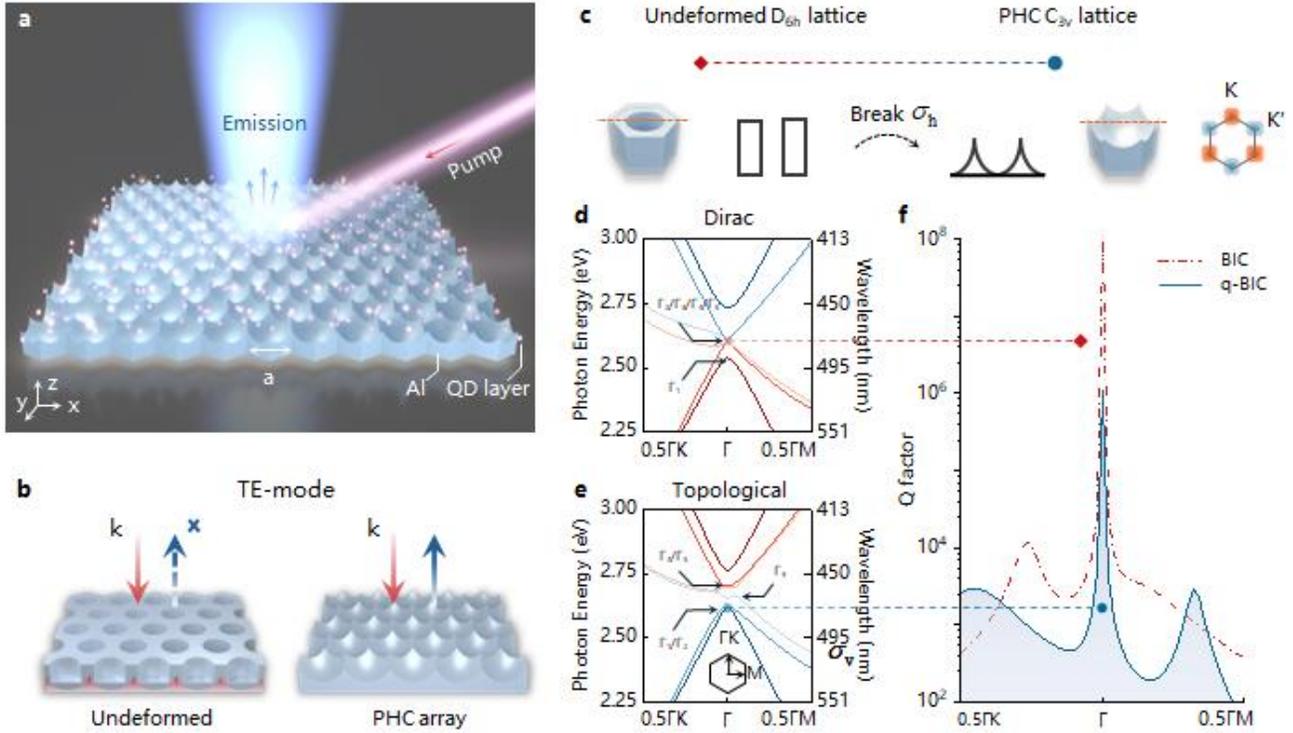

**Figure 1 $\sigma_h$ Symmetry-breaking Plasmonic Honeycomb Nanocavity**. **(a)** Schematic diagram of the studied structure: The $\sigma_h$ symmetry-broken q-BIC honeycomb nanocavity array interacts with a monolayer of pure blue-emitting perovskite quantum dots deposited on its surface. The plasmonic honeycomb nanocavities have a hexagonal lattice, forming nanoantenna at the K-point. **(b)** Schematic structures of prepared undeformed lattice arrays and PHC arrays with $\sigma_h$ symmetry breaking. The undeformed lattice array with TE polarization (left panel) reveals a $\sigma_h$ symmetry-broken aluminum nanocone layer after wetting chemical etching (right panel), transforming the ideal BIC into a q-BIC. Arrows indicate the open radiation channels. **(c)** Conceptual illustration of the $\sigma_h$ symmetry-breaking BIC. Using a wet chemical method, the $D_{6h}$ lattice on the anodic alumina template (left) is etched to create aluminum honeycomb nanocavity pores with $C_{3v}$ symmetry (right). The orange dashed line on the lattice indicates the location of the cross sections for both deformed and undeformed structures. The lattice with preserved $D_{6h}$ symmetry supports symmetry-protected BICs that cannot radiate energy into the far field under normal incidence ($\Gamma$ point). **(d, e)** Band structures of TE-like modes for the undeformed and honeycomb lattices shown in **(b)**, respectively. The band structure of the undeformed lattice shows a fourfold degenerate Dirac point at the $\Gamma$ point. When $\sigma_h$ symmetry breaking is introduced, the Dirac point of the honeycomb lattice reopens. **(f)** Q-factors corresponding to the $\Gamma_4$ band of the undeformed lattice and the $\Gamma_2$ band of the honeycomb lattice. The Q-factor of the undeformed lattice at the center of the Brillouin zone ($\Gamma$ point) approaches infinity, demonstrating symmetry-protected BIC properties. With the introduction of $\sigma_h$ symmetry breaking, the symmetry-protected BIC transitions to a q-BIC mode with finite Q ($10^6$).

The $\sigma_h$ symmetry-breaking plasmonic honeycomb nanocavity (PHC) array interacts with a deposited monolayer of perovskite quantum dots (PQDs, purchased from biotyscience, ABO-Cd-7-48) under optical excitation, as illustrated in Figure 1a. This structure is fabricated by forming a hexagonal lattice anodic aluminum oxide (AAO) template on pre-imprinted aluminum foil using guided anodization approach [45], followed by wet chemical etching to remove the alumina component (Figure 1b and supporting information 1). As a result, PHC comprises the exposed barrier layer after etching, creating a long-range ordered honeycomb nanocavity array in the xy-plane. This lithography-free nanofabrication technique has been widely applied in the production of nanoscale metasurfaces, flexible electronic devices, and other applications requiring precise patterning. It is crucial for realizing PHC arrays.

The $\sigma_h$ symmetry breaking is introduced after the wet chemical etching process, as illustrated in Figure 1c. The left side of the Figure 1c shows a six-fold symmetry ($D_{6h}$) triangular lattice belongs to the group $P6m \times \sigma_h$ with reflection symmetry $\sigma_h$ in the horizontal direction (xy-plane) [46], a cross-sectional view along the yz-plane is extracted at the orange dashed line, with a lattice constant (a = 400 nm) and an air hole radius ratio of 0.7*a. In accordance with the $D_{6h}$ symmetry, the ideal structure displays a fourfold degenerate Dirac point ($\Gamma_2/\Gamma_3/\Gamma_4/\Gamma_5$) at the $\Gamma$ point for TE-like polarization. The BICs dark modes ($\Gamma_4/\Gamma_5$) manifest as quadrupole modes, situated within the high-frequency band and obscured by dipole modes (Supporting Information 2), as illustrated in the band structure depicted in Figure 1d. The $D_{6h}$ symmetry of the lattice reduces to $C_{6v}$ due to $\sigma_h$ symmetry breaking, but neighboring lattice points are related to each other through the parity reversal symmetry (K, K'), thus the PHC lattice possesses $C_{3v}$ symmetry, as shown on the right side of Figure 1b. The honeycomb lattice gives rise to the formation of recessed nanocavities at the $\Gamma$ point, while tip nanocone are generated at the compressed regions between lattices (K-M). The breaking of the $\sigma_h$ symmetry leads to the opening of the Dirac point and the degeneracy of the dielectric ($\Gamma_1$) and air ($\Gamma_2$) bands at the $\Gamma$ point, as show in Figure 1e. The quadrupole modes ($\Gamma_1/\Gamma_2$) transiting to lower frequencies, while the dipole modes emerge at higher frequencies. The topological nature of these bands will be addressed subsequently. Prior to etching, the system possesses $D_{6h}$ symmetry, and the quadrupole modes ($\Gamma_4/\Gamma_5$) of the undeformed lattice manifest the BIC with an infinite quality factor (Q-factor), as illustrated by the red dashed line in Figure 1f .With the introduction of $\sigma_h$ symmetry breaking, the ideal BIC evolves into a q-BIC mode with a finite Q-factor ($10^6$), thereby creating a pathway for radiation, as indicated by the arrow in Figure 1b.

To experimentally characterize and verify the $\sigma_h$ symmetry-breaking q-BICs, a PHC array was fabricated with lattice constants a = 400 nm and depth h = 125 nm. Figure 2b depicts a scanning electron microscopy (SEM) image. Further characterization was conducted using an atomic force microscope (AFM) to reconstruct the three-dimensional (3D) morphology of the structure, as depicted in Figure 2c. This profile demonstrated a depression at the $\Gamma$ point of the lattice and the formation of nanowires at the K point. Height data extracted between next-nearest-neighbor (NNN) lattice (red dashed lines) and nearest-neighbor (NN) lattice (green dashed lines) are depicted in the upper and lower panels of Figure 2d, respectively. The mean height differential between nearest-neighbor unit cells, which do not traverse the apex of the nanowire, is 100 nm. Conversely, the cross-section between next-nearest-neighbor unit cells connects the extremities of two nanowires, resulting in an average height from the base of the nanocavity to the tip of the nanowire of 125 nm.

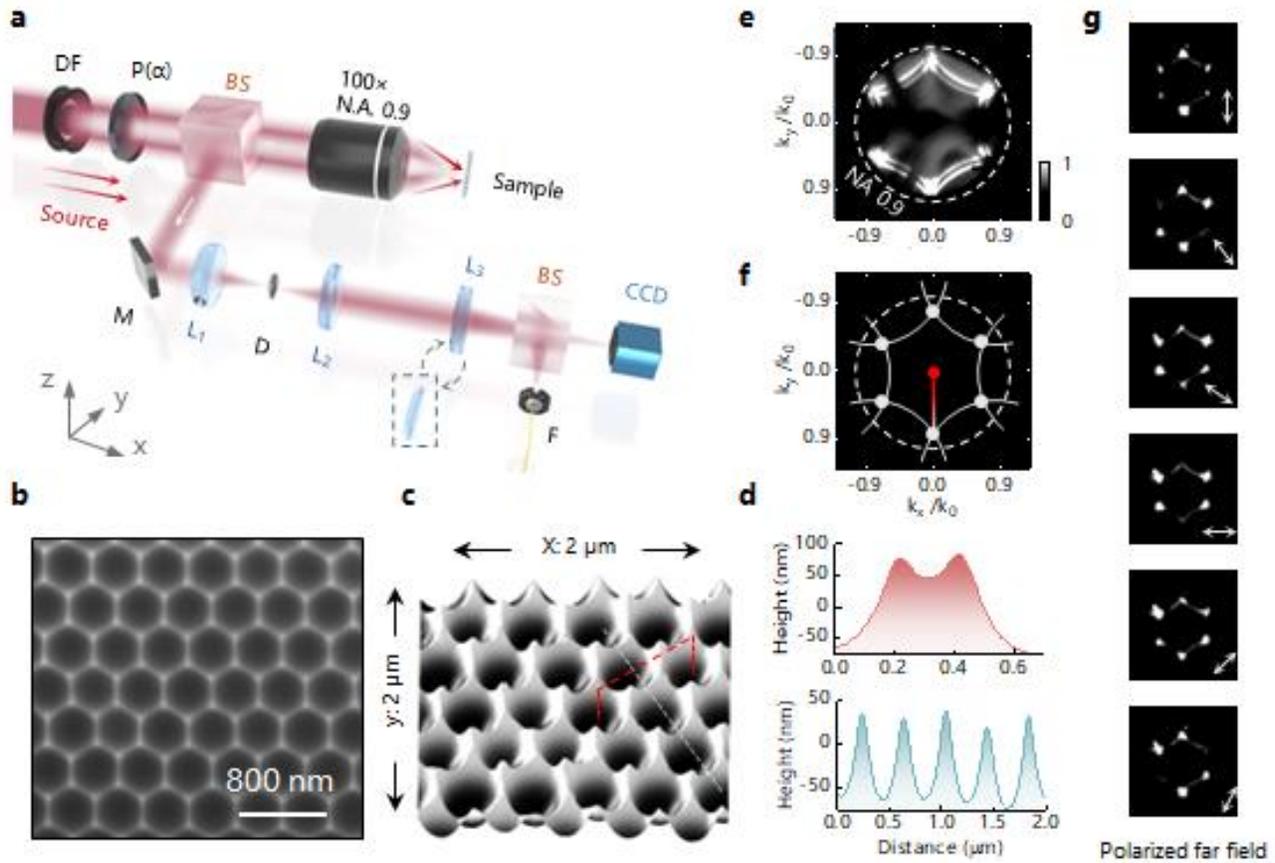

**Figure 2 Schematic diagram of the experimental setup characterizing the $\sigma_h$ symmetry breaking PHC array q-BICs (a)** Schematic diagram of the common optical path for angle-resolved resonance scattering and photoluminescence. DF: dark field: P: linear polarizer; BS: beam splitter; M: mirror; L: lens; PH: pinhole; F: filter. **(b)** Scanning electron microscopy (SEM) image of the PHC array. **(c)** Atomic force microscopy (AFM) image of the PHC array. **(d)** A height profile is presented between the next-nearest neighbor (NNN) lattice sites along the red dashed line in **(c)** (top image), and the height profile between the nearest neighbor (NN) lattice sites along the blue dashed line in **(c)**. **(e)** A Fourier back focal plane image of the PHC mercury lamp (546 & 577 nm) excitation reveals two arcs along the periodic boundary direction of the hexagonal lattice. The point of intersection of the arcs represents the high symmetry point, designated as the K point. **(f)** Schematic illustration of reciprocal space scanning for wave vector dispersion, with the spectral acquisition scan trajectory marked by a red line tracing the Γ-K point **(g)** Far-field polarization characteristics of the PHC array.

A Fourier back plane (FBP) imaging-based angular-resolved resonance scattering spectroscopy (ARSS) setup (Figure 2a) has been constructed for the purpose of characterizing the optical response of PHC both before and after the deposition of PQDs (see Methods for details). The sample was excited by a halogen lamp for the resonance scattering measurements. Incident light passes through a linear polarizer along the y-direction and then enters the beam-splitting prism. Further, a dark-field module was employed to project the incident light field onto an objective lens (×100, NA = 0.9), with a large-angle incidence onto the sample surface. The diffracted light from the sample was collected by the same objective lens and directed into a 4f system consisting of three lenses and a pinhole. Different focal length lenses were inserted in front of a CCD camera to allow easy switching between real and FBP images [47]. Finally, the Fourier plane of the diffracted light was projected via beam splitter BS2 onto CCD and fiber spectrometer. We used fiber with a core diameter of 100 μm to collect spectral signals displayed as spots on the Fourier plane (supporting 3). This configuration effectively eliminated scattered light interference from other positions around the sample, ensuring precise and focused light collection to produce sharp resonance peaks. To quickly locate high-symmetry points of the structure, two characteristic wavelengths of a mercury lamp (546 & 577 nm) were first used as the incident light source to image the first Brillouin zone boundary of the structure, as shown in Figure 2e. Two equifrequency contours were extracted along periodic boundary conditions, with intersections of curved equifrequency contours marking the position of the K point, thereby defining the K-Γ-M path. Figure 2f depicts a schematic of reciprocal space scanning for wave vector dispersion, with the K point marked by a blue dot and spectral acquisition paths indicated by red arrows. Simultaneously, a 30° rotational step was applied to a rotating polarizer to determine the far-field polarization state for each momentum state (Figure 2g).

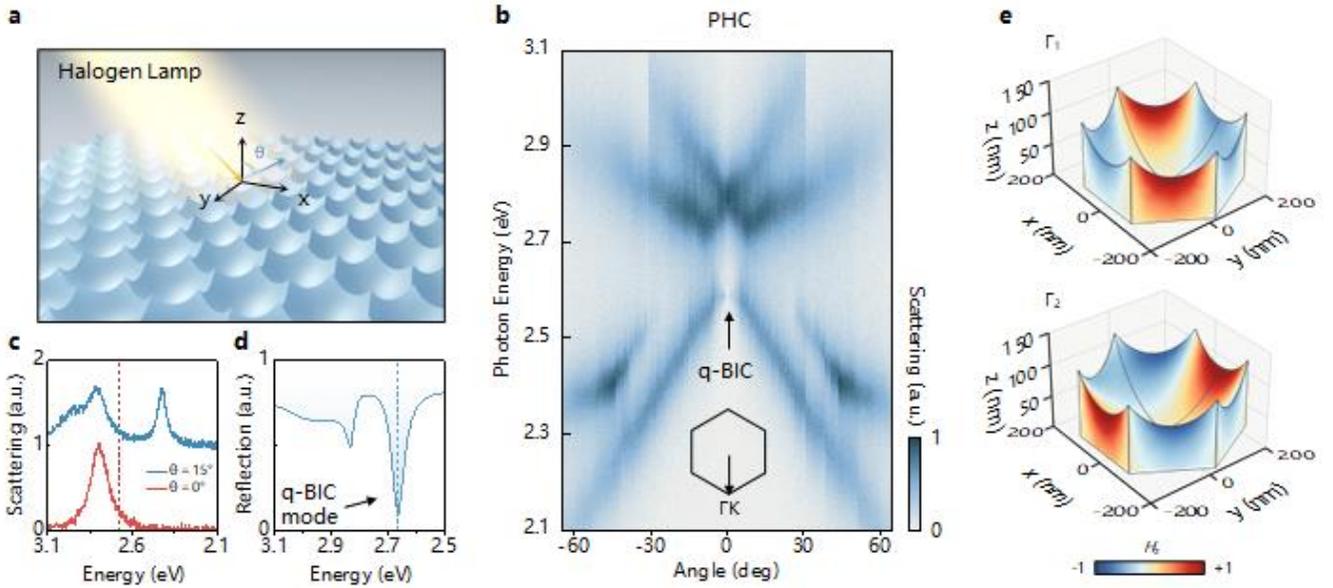

**Figure 3 $\sigma_h$ symmetric-breaking q-BICs features.** (**a**) Schematic illustrating ARSS measurements, where the halogen lamp beam is directed onto the sample through a dark-field objective lens. (**b**) Experimental TE-like angle resolved resonance scattering spectrum of the PHC array structure, with the scanning path at Γ-K, where $\theta$ in the figure represents the collection angle. The disappearance of scattering signal at Γ in the low-frequency band indicates the presence of a q-BIC. (a.u., arbitrary units). (**c**) ARSS extracted from (b) at collection angles of 0° and 15°. (**d**) Computed reflectance spectrum of the PHC at Γ point, showing a reflectance dip at 2.638 eV suggestive of a q-BIC. (**e**) Three-dimensional mode analysis revealing the degeneracy of two modes of the PHC lattice at the Γ point, identified as quadrupole modes (normalized electric field, $H_z$ component), characterized by their symmetry decoupled from plane wave mode solutions.

The incident light source was switched to halogen lamp, and using the above method, angular-resolved resonance scattering data were collected for the PHC, yielding [$\lambda$, $\sin\theta$], which was converted to EN using the formula $EN = \hbar c / \lambda$. The PHC was excited under different wavelengths of incident light, inducing resonances of the guided modes of the periodic structure, which were ultimately collected by the objective lens as scattering. Angle-resolved resonance scattering measurements were advantageous in eliminating Fabry-Perot modulation signals inherent to the material itself. Figure 3b shows the TE-like dispersion band diagram obtained from the measurements, revealing degenerate dipole emissions at 2.755 eV. Notably, the localized disappearance of resonance scattering at $\theta = 0$ and 2.61 eV confirms the decoupling of resonance modes from the radiative electromagnetic continuum as expected (Figure 1d). The q-BICs of the PHC, designed successfully at blue-light (475 nm / 2.61 eV) in the $\Gamma_1 / \Gamma_2$. Spectra extracted from Figure 3b at $\theta = 0°$ and 15° show the disappearance of the first-order scattering resonance at the $\Gamma$ point (long wavelength), as shown in Figure 3c, and the red dashed line indicates the resonance position of q-BIC. A PHC lattice with a = 400 nm and h = 125 nm was constructed in COMSOL, and the reflection spectrum of this lattice was calculated (figure 3d), revealing a reflection valley at 2.638 eV at positive incidence ($\Gamma$ point).

To further ascertain the robust presence of q-BICs in PHC after introducing $\sigma_h$ symmetry breaking, three-dimensional mode analysis identified degenerate quadrupole modes at 2.638 eV ($\Gamma_1 / \Gamma_2$). The TE-like electric field patterns of these two q-BICs are shown in Figure 3e, with normalized magnetic field $H_z$ components evident both externally and internally within the lattice. The distribution of $H_z$ magnetic field modes on the vertical plane (xy plane) under $C_2$ symmetry exhibits even parity [18,39], confirming the robustness of these symmetry-protected BICs under $\sigma_h$ symmetry breaking.

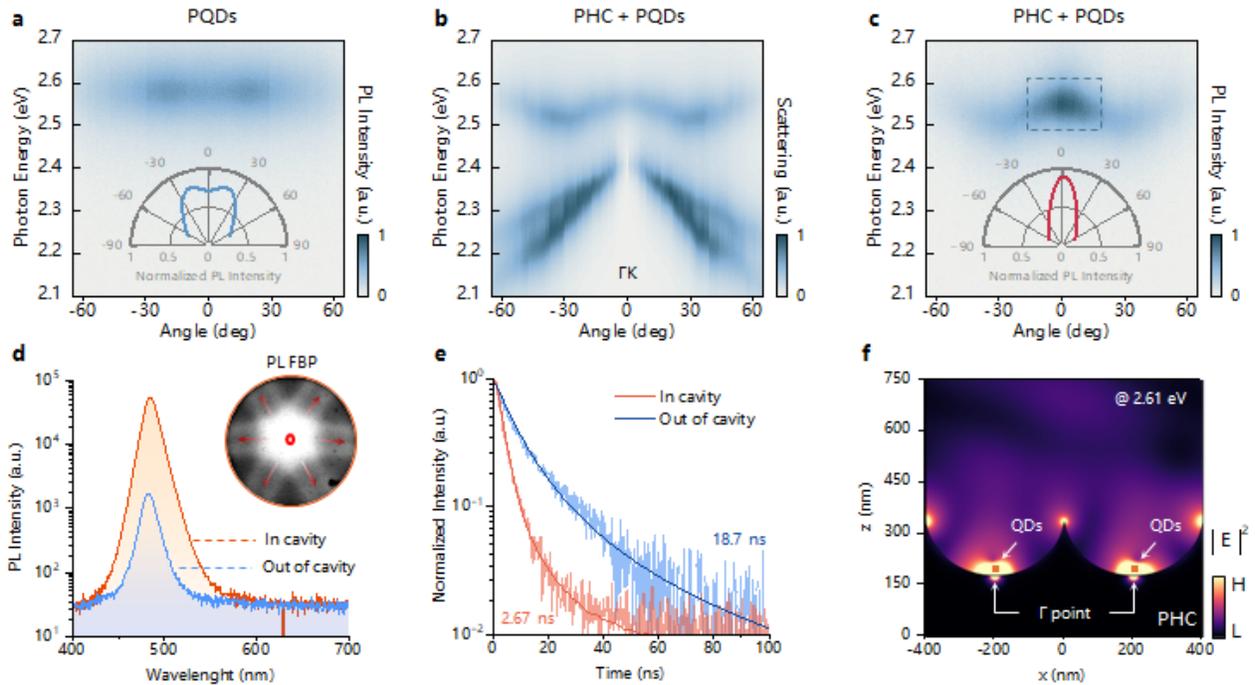

**Figure 4 Purcell enhancement effect of the q-BICs PHC.** (a) Experimental results of the angular-resolved photoluminescence (ARPL) of PQDs deposited on pure aluminum foil. The inset depicts the angular-resolved intensity distribution contour, which evinces a dipole distribution of PQDs emission at 2.605 eV. (b) Experimental TE-like angle resolved resonance scattering spectrum of a single layer of PQDs deposited on a PHC array. (c) ARPL results of PQDs

coupled to PHC. The inset shows the angular-resolved intensity distribution contour, demonstrating vertical emission of the PHC resonant cavity at 2.583 eV, with emission angles concentrated from ±30° to ±5°. **(d)** Normal incidence photoluminescence spectra of PQDs in the PHC array region (red curve) and in the flat region (blue curve). The inset displays the PL-FBP image, where PQDs emission is concentrated at the Γ point and rapidly decays along high-symmetry directions, with the edge emissions modulated into Bloch-type patterns. **(e)** PL decay curves measured for PQDs coupled to the PHC array structure and the flat region. **(f)** The EM - field intensity of the PQD located at the Γ-point of the PHC lattice at the periodic boundary under q-BIC resonance wavelength (2.61 eV) excitation has been calculated for fixed parameters a = 400 nm and h = 125 nm. H, high; L, low.

We proceed to discuss the enhancement and modification of exciton emission using $\sigma_h$ symmetry-breaking q-BICs. In this study, fully inorganic cesium lead halide perovskite quantum dots ($CsPbX_3$, X is mixed halide systems (Cl/Br)) were specifically chosen as the exciton material for the plasmonic cavity-emitter system. At resonance, the coupling between the plasmonic cavity and excitons results in an enhanced radiative recombination rate, effectively reducing the exciton radiative lifetime. The degree of radiative enhancement is quantified by the Purcell factor.

$$F = \frac{3}{4\pi^2} \frac{Q}{V_{\text{eff}}} \left(\frac{\lambda}{2n}\right)^3, \qquad (2)$$

where Q is the quality factor of the plasmonic cavity, $V_{\text{eff}}$ is the effective mode volume, $\lambda$ is the emission wavelength, and *n* is the refractive index of the medium.

Before coupling the PQDs with the PHC, we set the incident light to x-polarization as illustrated in Figure 2a, then excited the PQDs covered aluminum foil (flat region) with 366 nm light to measure the TE-like ARPL spectrum (Figure 4a). The experimental data show that the emission peaks along the K-Γ path are all located at 2.61 eV, which spectrally overlaps with the q-BIC resonance mode of the PHC. Additionally, the blue curve in the inset represents the fitted angular-resolved intensity profile, indicating that emission from the PQDs outside the cavity is primarily directed with maximum intensity toward 30°. This suggests that the out-of-cavity radiative mode has a distinct dipole (low Q-factor) emission pattern.

Next, we spin-coated PQDs dispersed in toluene (10 mg/mL for 60 s at 1500 rpm) onto the PHC substrate. Because the PHC backbone avoids vertical air holes, it was easy to coat a monolayer of PQDs film on the PHC (SEM images can be seen in Supporting 1). Similar to Figure 3b, we experimentally investigated the plasmonic cavity-emitter system under halogen lamp illumination using ARSS, the experimental results are shown in Figure 4b. Along the high-symmetry direction K-Γ-K, the characteristic spectrum of PQDs appears around 2.583 eV. Compared with Figure 4a, which does not consider the coupling to the PHC, the scattering band of the PQDs deposited on the PHC shows an obvious Bloch modulation, confirming the coupling between the plasmonic cavity and the emitter. After coupling, the symmetry of the polariton resonance mode in the plasmonic cavity-emitter system ensures the presence of a q-BIC, resulting in a new q-BIC resonance scattering at 2.384 eV. It is important to emphasize that the field enhancement provided by the original PHC q-BIC (2.61 eV) does not disappear due to coupling nor with the emergence of a new q-BIC.

To investigate the modulation of the emitter emission characteristics by the $\sigma_h$ symmetry-broken plasmonic lattice cavity, we have measured the ARPL spectrum of the coupled PHC-PQDs system. The TE-like normalized photoluminescence dispersion is shown in Figure 4c. Under ultraviolet pumping at 366 nm, we collected the emission band of the PQDs along the high-symmetry direction K-Γ-K. The experimental data show a prominent PQD emission peak at 2.562 eV at the Γ point. In contrast to the emission from the PQDs outside the cavity (Figure 4a), away from the Γ point, the PQD emission inside the cavity shows a rapid decay indicating that the q-BIC is no longer protected by the $C_6$ symmetry, resulting in a significant decrease in Q-factor. Inset shows the fitted angular-resolved intensity profile at 2.562 eV. The red curve in the inset represents the angular-resolved intensity distribution profile at 2.562 eV. It illustrates the generation of vertically oriented emission from the cavity after coupling, with the emission beam's full width at half maximum measuring 12.6°. Notably, when the collection angle is adjusted, the far-field emission of the PQDs coupled to the q-BIC cavity exhibits a red shift, with the angle-tuning range from 2.52 to 2.594 eV. This demonstrates that the emission peak of the PQDs can be effectively tuned by coupling with the external cavity after chemical synthesis. Since the emission linewidth of the PQDs does not cover the low-frequency resonance of the plasmonic cavity polaritons, the dispersion band is not populated. To verify the Purcell enhancement provided by the PHC cavity to the PQDs, we compared the ARPL emission intensities of PQDs coupled inside the PHC cavity with those of the same concentration of PQDs outside the cavity (i.e., the flat region of the substrate). Their PL spectra were collected at a reflection collection angle of 0°, as shown in Figure 4d. The inset shows that under 366 nm excitation, the far-field emission from the PQDs inside the cavity becomes more concentrated, exhibiting a honeycomb lattice plane profile. In addition, a small portion of the emission is observed along the high-symmetry directions of the lattice. This directional far-field emission characteristic can be attributed to the lattice symmetry. In the hexagonal lattice, TE-like modes yield six in-plane propagating Bloch waves and one out-of-plane propagating Bloch wave, analogous to those observed in vertical cavity surface-emitting lasers [48].

    The plasmonic cavity significantly enhances the exciton luminescence. We measured the PL lifetime of PQDs both inside and outside the PHC cavity at the same concentration, as shown in Figure 4e. Using a 405 nm x-polarized picosecond pulse laser for excitation at room temperature, we determined the PL decay rates by fitting the experimental data to an exponential decay function. The PL decay rate of PQDs in the flat region is 18.7 ns, close to the lifetime of the particles in toluene, which is attributed to the combined effects of radiative and non-radiative exciton recombination mechanisms at this temperature. In contrast, the PL lifetime of PQDs coupled to the PHC cavity is reduced to 2.67 ns under identical pumping conditions. This shorter PL lifetime indicates an effective enhancement of the radiative exciton recombination rate and the excitation rate by the PHC cavity. The design of the $\sigma_h$-symmetry-broken PHC cavity retains plasmonic tips that support very small mode volumes. To investigate the Purcell effect at these nanocavity, we used COMSOL simulations to evaluate the near-field enhancement contribution of PQDs placed at the Γ of the PHC lattice, as shown in Figure 4f. The point dipole sources were placed in the PHC cavity to simulate quantum dot emission (with x-polarized emission), and TE-polarized incident light along the y-direction (normal incidence) excited the PHC-PQD. The PQD located at the Γ point of the PHC lattice exhibit significant near-field enhancement under q-BIC resonance wavelength excitation. Furthermore, in addition to the q-BIC cavity field enhancement provided by the PHC itself, the LSPR at the nanocone (K-point) provides a secondary enhancement of the emission of the PQDs to the far field [45] (Supporting Material Figure S4). We also compared the enhancement effects of the un-deformed ($D_{6h}$)

cavity and the PHC ($C_{3v}$) with broken $\sigma_h$ symmetry to the PQD and the results showed a more significant enhancement emission by the PHC. Meanwhile, we calculated the time-averaged power flow of the excited emission of PQDs in both cavities, the Poynting vectors ($\mathbf{S} = \mathbf{E} \times \mathbf{H}$) show that the emission direction of PQDs within the PHC is vertical to the cavity surface, which is consistent with the emission direction shown in Figure 1b. In contrast, the emission of PQDs inside an un-deformed cavity exhibits only transversal driving. Therefore, the PHC with broken $\sigma_h$ symmetry is able to provide an out-of-plane wavevector for the emission of PQDs, driving the collective emission of PQDs to the far field along a direction perpendicular to the cavity surface. The experimental and simulation data confirm that the PHC can effectively enhance the directional emission and luminescence intensity of PQDs, providing a non-invasive method to change the emission characteristics of the proton cavity.

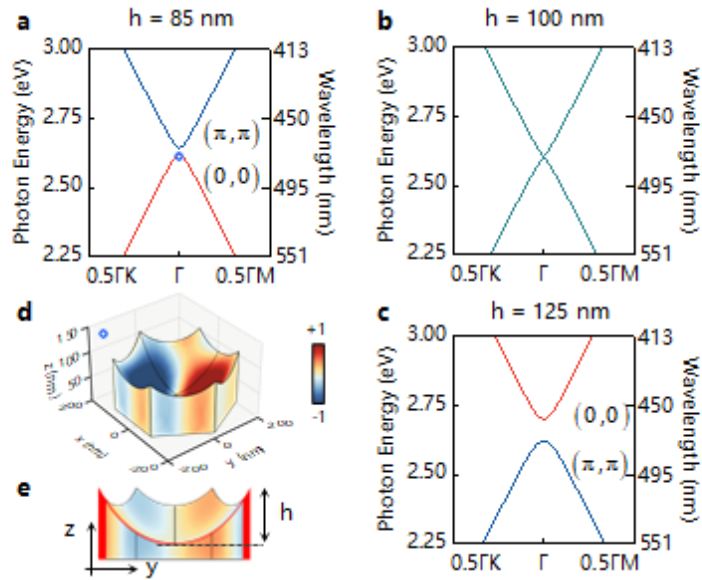

**Figure 5 Topological phase transition induced by Zak phase evolution. (a, b, c)** corresponds to the evolution of the Zak phase with height $h$. The band gap is closed at $h = 100$ nm and the energy bands are flipped after this transition point. **(d)** Mode field distribution $H_z$ of the bulk mode at the Γ point in the low-frequency band for $h = 85$ nm. The xy-plane of the 3D mode distribution is shown as a singular mode under $C_2$, coupled to a plane wave. **(e)** Profile of a honeycomb nanocavity protocell, $h$ is the distance from the tip of point k to the bottom of the hole.

In the design of $\sigma_h$ symmetry-breaking PHC, we noticed that two-dimensional Dirac points on the energy bands suffer from a closing-to-opening process (Figures 1c,d), similar to the topological engineering of band structure in one-dimensional dielectric PCs [49]. Consequently, we theoretically investigated the topological properties of the band structure of the PHC system. Indeed, when the lattice constant of the plasmonic cavity is the same order of magnitude as the resonance wavelength, the incident light in the lattice generates Bloch waves (collective diffraction modes) propagating in the facets, which can be strongly coupled as a discrete state to the PLR in the metallic nanostructures to produce plasmonic lattice resonances. As a direct result of this strong coupling, Rabi splitting occurs at the 2D Dirac point, with the band gap size calculable from coupled mode theory. The bright and q-BIC modes are thus separated. Upon successive variations of the system parameters, we observe a topological energy band flip, corresponding to a shift in the geometrical phase of the hybrid energy band. This $\sigma_h$ symmetry-breaking topological phase transition can be characterized by the 2D Zak phase, defined as follows:

$$\theta_\alpha = 2\pi \mathbf{P}_\alpha = \frac{1}{2\pi}\int_{FBZ} \text{Tr}[A_\alpha(\mathbf{k})]d\mathbf{k}, \quad \alpha = x, y, \tag{3}$$

where $\mathbf{P} = [Px, Py]$ indicates the 2D polarization of the lattice, $A_a(\mathbf{k}) = i\langle \psi_{n,k} | \partial_{k_a} | \psi_{n,k} \rangle$ is the Berry connection, and $|\psi_k\rangle$ denotes the Bloch eigenfunction on the nth band with wave vector $k$. We selected the Γ point of the PHC lattice as the inversion center and calculated the 2D Zak phase of $\Gamma_2/\Gamma_5$, as shown in Figures 5a-c. When the depth $h = 85$ *nm*, the lower and upper bulk band are characterized by the Zak phases of (0,0) and (π, π), respectively. In contrast, a clear Zak phase transition is observed when $h = 105$ *nm* (Figure 5c), with the lower and upper band having (π, π) and (0, 0) Zak phases, respectively, confirming the band inversion. It is also noted that the band gap closes when $h = 100$ *nm*, and the Zak phase cannot be defined (Figure 5b). Also, this topological phase transition can also be determined by the parity inversion under $C_2$ symmetry. As shown in Figure 5d, the 3D mode distribution ($h = 85$ *nm*) of the $\Gamma_2$ mode reveals that the $H_z$ field in the xy-plane exhibits odd parity. Whereas the mode profile at $\Gamma_2$ has an even parity when $h$ increases to 125 *nm* (Figure 3e). Noting that the q-BIC of the PHC lattice always appears in the band with $\theta_\alpha = \pi$, the band inversion suggests that the in-plane Bloch wave of the PHC lattice accumulates an additional phase under continuous variations of the system parameters. This is a consequence of the strong coupling between the Bloch wave with the LSPR to produce plasmonic lattice resonances. Therefore, this geometrical phase can be realized by modulating the lattice hole depth (nanoantenna height $h$) such that the LSPRs are strongly coupled to the in-plane Bloch waves. The 2D Zak phase can explicitly characterize the topological properties of $\sigma_h$ symmetry-breaking lattice, since the band structure and the parity inversion of the $\sigma_h$ symmetry-breaking (2.5D) lattice are similar to the 2D lattice, which is promising for the design of robust q-BIC resonant cavities. Meanwhile, the tuning mechanism of the geometrical phase opens up new ideas for generating robust on-chip topological plasmon photonic devices.

**Conclusion**

We demonstrate an integrated plasmonic nanocavity-quantum emitter system that supports robust q-BICs and efficiently couples with perovskite quantum dots on the cavity surface. This system realizes vertical emission in the blue region with a hexagonal beam shape ($\theta_{fwhm} = 12°$). Additionally, the breaking of $\sigma_h$ symmetry induces high-Q q-BICs in the PHC, while PLR in the K-M path further enhance the spontaneous emission intensity of the PQDs. The PHC supported q-BIC, with topologically non-trivial phases, extends the immunity of light-emitting devices to external perturbations. Thus, by utilizing $\sigma_h$ symmetry-broken PHCs, we achieve enhanced spontaneous emission from PQDs and effective beam tuning. Our proposed strategy to modulate the spontaneous radiation of PQDs using $\sigma_h$ symmetry-breaking PHCs may provide new design ideas for on-chip nanoscale optical manipulation applications, including high-efficiency lasers, harmonic modulation, and metalens optical imaging.

**Methods**
**Characterization.** Scanning electron microscope (MIRA3-LMH) in-beam SE mode was used to photograph the surface features of the samples at 20 kV. Samples were imaged using a material-based AFM (BRUKER, Dimension FastScan) in contact mode (scan rate). The AFM images were evaluated and analyzed using NanoScope Analysis 3.00, which revealed a clear 2 μm × 2 μm three-dimensional profile.
**Optical characterization (ARSS, ARPL, PL lifetime).** Angle-resolved resonance scattering spectroscopy (ARSS) and photoluminescence spectroscopy (ARPL) were both based on a Fourier back-plane (FBP) imaging device (Fig. 2a). For ARSS measurements, the sample was excited using a halogen lamp as the incident light source. A rotatable adjustable mount with a linear polarizer was used to adjust the direction of polarization of the incident light along the y-direction before it entered the beam-splitting prism. A dark field module (model) was used to project the incident light field onto a 100× Olympus objective that was corrected for high numerical aperture (NA = 0.9) and infinite conjugate and incident at a large angle to the sample surface. The diffracted light from the sample was collected by the same objective lens and directed into a 4f optical system, comprising three lenses and a pinhole. Finally, the Fourier plane of the diffracted light was projected via beam splitter BS2 onto CCD (QImaging Retiga R1 from Cairn Research) and fiber spectrometer (Exemplar®Plus, BTC655N). We used xx fiber with a core diameter of 100 μm (model) to collect spectral signals displayed as spots on the Fourier plane. The collection end of the fiber was fixed on a support rod connected to an electric displacement platform, enabling point-by-point scanning with a 0.1 mm step using an externally- drive system to achieve dispersion imaging of the sample in momentum space. This configuration effectively eliminated scattered light interference from other positions around the sample, ensuring precise and focused light collection to produce sharp resonance peaks. The same optical setup was employed for ARPL measurements, utilizing a mercury lamp ($\lambda$ = 366 nm) to vertically excite PQDs. After the PL signal from the sample was passed through a 400 nm long-pass filter, the Fourier image and ARPL spectrum of the signal were imaged simultaneously using the FBP device. For the time-resolved PL measurements, we obtained the fluorescent-emission decays of the samples using a
FLS1000 Photoluminescence Spectrometer. A picosecond pulsed diode laser ($\lambda$ = 405 nm) with pulse duration of 70 ps was used to excite the sample. The spectrally integrated PL in the range of 390 nm-700 nm were collected using a photomultiplier tube.

**Numerical Modeling.** The optical properties of the PHC array structure and the emission enhancement of the PQDs on its surface were modeled using the commercially available software COMSOL Multiphysics. The 3D model of the PHC lattice was obtained by Boolean operations on a hexagonal prism and an ellipsoid (long axis a, short axis b), where the center of the ellipsoid is highly aligned with the surface on the hexagonal prism. The 3D energy band structure of the PHC lattice and the Q-factors of the corresponding strips were calculated using eigenfrequency studies with periodic boundary conditions, and redundant pseudo-modes were manually eliminated. The 3D mode analysis and reflection spectra were calculated for this periodic structure using frequency domain studies. And PL enhancement was performed in a finite-size PHC array using a point electric dipole as a light source. The electric field distribution at the resonance wavelength was extracted by placing the point electric dipoles at the Γ and K points of the lattice, respectively, under the frequency domain study. Similarly, PL far-field emission was calculated in a 10 × 10 hexagonal PHC array.

**Supporting information**
PHC fabrication, SEM image of PQDs on PHC, un-deformed and PHC lattice Γ-point mode, Fourier back plane image of PHC, Comparison of $D_{6h}$ and $C_{3v}$ cavities for near-field enhancement of PQDs and directional driving of emission are included in the supporting information at https://xxx.xxx.

**Acknowledgements**
The authors thank the supporting of the National Natural Science Foundation of China (Grant No. 12274054, 12074054, 61905051).

**Author contribution**

Y.F. conceived the idea and directed the project. Y.C. performed the main experiment and the main FEM calculations. J.L and Y.G measured part of the FBP. W.W. directed the nanocavity fabrication. J.H. and Y.W. fabricated the nanocavity. B.D. directed the quantum dots synthesis. Y.M fabricated the quantum dots. Z.S performed the AFM measurement. H.W. performed the QD deposition. H.L performed part of the calculations. Y.C. and Y.F analyzed the data and wrote the manuscript. All of the authors discussed the data and revised the manuscript.

**Conflicts of interest**

The authors declare no competing financial interest.

**References**
[1]  J. Jang, M. Jeong, J. Lee, S. Kim, H. Yun, and J. Rho, Planar Optical Cavities Hybridized with Low-Dimensional Light-Emitting Materials, Adv. Mater. **35**, 2203889 (2023).
[2]  G. M. Akselrod, C. Argyropoulos, T. B. Hoang, C. Ciracì, C. Fang, J. Huang, D. R. Smith, and M. H. Mikkelsen, Probing the mechanisms of large Purcell enhancement in plasmonic nanoantennas, Nat. Photonics **8**, 11 (2014).
[3]  H. Chen et al., Sub-50-ns ultrafast upconversion luminescence of a rare-earth-doped nanoparticle, Nat. Photonics **16**, 9 (2022).
[4]  B.-Y. Wen et al., Manipulating the light-matter interactions in plasmonic nanocavities at 1 nm spatial resolution, Light Sci. Appl. **11**, 1 (2022).
[5]  A. Aigner, A. Tittl, J. Wang, T. Weber, Y. Kivshar, S. A. Maier, and H. Ren, Plasmonic bound states in the continuum to tailor light-matter coupling, Sci. Adv. **8**, eadd4816 (2022).
[6]  M. Jeong, B. Ko, C. Jung, J. Kim, J. Jang, J. Mun, J. Lee, S. Yun, S. Kim, and J. Rho, Printable Light-Emitting Metasurfaces with Enhanced Directional Photoluminescence, Nano Lett. **24**, 5783 (2024).
[7]  L. M. Berger, M. Barkey, S. A. Maier, and A. Tittl, Metallic and All-Dielectric Metasurfaces Sustaining Displacement-Mediated Bound States in the Continuum, Adv. Opt. Mater. **12**, 2301269 (2024).
[8]  K. A. Sergeeva et al., Laser-Printed Plasmonic Metasurface Supporting Bound States in the Continuum Enhances and Shapes Infrared Spontaneous Emission of Coupled HgTe Quantum Dots, Adv. Funct. Mater. **33**, 2307660 (2023).
[9]  N. Parappurath, F. Alpeggiani, L. Kuipers, and E. Verhagen, Direct observation of topological edge states in silicon photonic crystals: Spin, dispersion, and chiral routing, Sci. Adv. **6**, eaaw4137 (2020).
[10] R. Barczyk, N. Parappurath, S. Arora, T. Bauer, L. Kuipers, and E. Verhagen, Interplay of Leakage Radiation and Protection in Topological Photonic Crystal Cavities, Laser Photonics Rev. **16**, 2200071 (2022).
[11] L. Yang, G. Li, X. Gao, and L. Lu, Topological-cavity surface-emitting laser, Nat. Photonics **16**, 4 (2022).
[12] A. Dikopoltsev et al., Topological insulator vertical-cavity laser array, Science **373**, 6562 (2021).
[13] Y. Yang, Y. Yamagami, X. Yu, P. Pitchappa, J. Webber, B. Zhang, M. Fujita, T. Nagatsuma, and R. Singh, Terahertz topological photonics for on-chip communication, Nat. Photonics **14**, (2020).
[14] Y. Chen, Chiral topological whispering gallery modes formed by gyromagnetic photonic crystals, Phys. Rev. B (2023).
[15] R. de Oliveira, M. Colombano, F. Malabat, M. Morassi, A. Lemaître, and I. Favero, Whispering-Gallery Quantum-Well Exciton Polaritons in an Indium Gallium Arsenide Microdisk Cavity, Phys. Rev. Lett. **132**, 126901 (2024).
[16] *Single-Molecule Optofluidic Microsensor with Interface Whispering Gallery Modes*, https://doi.org/10.1073/pnas.2108678119.
[17] H. Lee, T.-Y. Lee, Y. Park, K.-S. Cho, Y.-G. Rho, H. Choo, and H. Jeon, Structurally engineered colloidal quantum dot phosphor using TiO2 photonic crystal backbone, Light Sci. Appl. **11**, 1 (2022).
[18] S. Han et al., Electrically-pumped compact topological bulk lasers driven by band-inverted bound states in the continuum, Light Sci. Appl. **12**, 145 (2023).

[19] M. Rao, F. Shi, Z. Rao, J. Yang, C. Song, X. Chen, J. Dong, Y. Yu, and S. Yu, Single photon emitter deterministically coupled to a topological corner state, Light Sci. Appl. **13**, 1 (2024).
[20] Z. Chen et al., Deep learning with coherent VCSEL neural networks, Nat. Photonics (2023).
[21] L. Kühner, L. Sortino, R. Berté, J. Wang, H. Ren, S. A. Maier, Y. Kivshar, and A. Tittl, Radial bound states in the continuum for polarization-invariant nanophotonics, Nat. Commun. **13**, 4992 (2022).
[22] F. Yesilkoy, E. R. Arvelo, Y. Jahani, M. Liu, A. Tittl, V. Cevher, Y. Kivshar, and H. Altug, Ultrasensitive hyperspectral imaging and biodetection enabled by dielectric metasurfaces, Nat. Photonics **13**, 390 (2019).
[23] O. A. M. Abdelraouf, A. P. Anthur, X. R. Wang, Q. J. Wang, and H. Liu, Modal Phase-Matched Bound States in the Continuum for Enhancing Third Harmonic Generation of Deep Ultraviolet Emission, ACS Nano **18**, 4388 (2024).
[24] L. Valencia Molina, R. Camacho Morales, J. Zhang, R. Schiek, I. Staude, A. A. Sukhorukov, and D. N. Neshev, Enhanced Infrared Vision by Nonlinear Up-Conversion in Nonlocal Metasurfaces, Adv. Mater. **36**, 2402777 (2024).
[25] Z.-K. Shao, H.-Z. Chen, S. Wang, X.-R. Mao, Z.-Q. Yang, S.-L. Wang, X.-X. Wang, X. Hu, and R.-M. Ma, A high-performance topological bulk laser based on band-inversion-induced reflection, Nat. Nanotechnol. **15**, 1 (2020).
[26] T. B. Hoang, G. M. Akselrod, A. Yang, T. W. Odom, and M. H. Mikkelsen, Millimeter-Scale Spatial Coherence from a Plasmon Laser, Nano Lett. **17**, 6690 (2017).
[27] Y. Fang, N. Gao, and L. Shao, Photoemission Enhancement of Plasmonic Hot Electrons by Au Antenna–Sensitizer Complexes, ACS Nano (2024).
[28] J. Pettine et al., Light-driven nanoscale vectorial currents, Nature **626**, 8001 (2024).
[29] W. Wang, L. V. Besteiro, P. Yu, F. Lin, A. O. Govorov, H. Xu, and Z. Wang, Plasmonic hot-electron photodetection with quasi-bound states in the continuum and guided resonances, Nanophotonics **10**, 1911 (2021).
[30] Y. Xu, H. Hu, W. Chen, P. Suo, Y. Zhang, S. Zhang, and H. Xu, Phononic Cavity Optomechanics of Atomically Thin Crystal in Plasmonic Nanocavity, ACS Nano **16**, 12711 (2022).
[31] H. Hu, A. K. Pal, A. Berestennikov, T. Weber, A. Stefancu, E. Cortés, S. A. Maier, and A. Tittl, Surface-Enhanced Raman Scattering in BIC-Driven Semiconductor Metasurfaces, Adv. Opt. Mater. **n/a**, 2302812 (n.d.).
[32] R. Gao, Y. He, D. Zhang, G. Sun, J.-X. He, J.-F. Li, M.-D. Li, and Z. Yang, Gigahertz optoacoustic vibration in Sub-5 nm tip-supported nano-optomechanical metasurface, Nat. Commun. **14**, 485 (2023).
[33] A. Yu. Bykov, Y. Xie, A. V. Krasavin, and A. V. Zayats, Broadband Transient Response and Wavelength-Tunable Photoacoustics in Plasmonic Hetero-nanoparticles, Nano Lett. **23**, 2786 (2023).
[34] V. G. Kravets, A. V. Kabashin, W. L. Barnes, and A. N. Grigorenko, Plasmonic Surface Lattice Resonances: A Review of Properties and Applications, Chem. Rev. **118**, 5912 (2018).
[35] Y. Liang, D. P. Tsai, and Y. Kivshar, From Local to Nonlocal High-$Q$ Plasmonic Metasurfaces, Phys. Rev. Lett. **133**, 053801 (2024).
[36] Y. Zhang, B. Zhang, Z. Li, M. Feng, H. Ling, X. Zhang, X. Wang, Q. Wang, A. Song, and H.-T. Chen, Surface plasmon-cavity hybrid state and its graphene modulation at THz frequencies, Nanophotonics **13**, 2207 (2024).
[37] K. V. Sreekanth, J. Perumal, U. S. Dinish, P. Prabhathan, Y. Liu, R. Singh, M. Olivo, and J. Teng, Tunable Tamm plasmon cavity as a scalable biosensing platform for surface enhanced resonance Raman spectroscopy, Nat. Commun. **14**, 7085 (2023).
[38] N. S. Mueller et al., Photoluminescence upconversion in monolayer WSe2 activated by plasmonic cavities through resonant excitation of dark excitons, Nat. Commun. **14**, 1 (2023).
[39] C. W. Hsu, B. Zhen, A. D. Stone, J. D. Joannopoulos, and M. Soljačić, Bound states in the continuum, Nat. Rev. Mater. **1**, 9 (2016).
[40] Y. Liang, K. Koshelev, F. Zhang, H. Lin, S. Lin, J. Wu, B. Jia, and Y. Kivshar, Bound States in the Continuum in Anisotropic Plasmonic Metasurfaces, Nano Lett. **20**, 6351 (2020).
[41] N. H. M. Dang et al., Realization of Polaritonic Topological Charge at Room Temperature Using Polariton Bound States in the Continuum from Perovskite Metasurface, Adv. Opt. Mater. **10**, 2102386 (2022).
[42] C. Schiattarella, Directive giant upconversion by supercritical bound states in the continuum, (n.d.).
15

[43] Y. Chen et al., Observation of intrinsic chiral bound states in the continuum, Nature **613**, 474 (2023).
[44] E. Csányi et al., Engineering and Controlling Perovskite Emissions via Optical Quasi-Bound-States-in-the-Continuum, Adv. Funct. Mater. **34**, 2309539 (2024).
[45] F. Lv, J. La, S. He, Y. Liu, Y. Huang, Y. Wang, and W. Wang, Off-Angle Amplified Spontaneous Emission of Upconversion Nanoparticles by Propagating Lattice Plasmons, ACS Appl. Mater. Interfaces **14**, 54304 (2022).
[46] R. Guo, M. Nečada, T. K. Hakala, A. I. Väkeväinen, and P. Törmä, Lasing at K Points of a Honeycomb Plasmonic Lattice, Phys. Rev. Lett. **122**, 013901 (2019).
[47] Y. Guo, J. Li, and Y. Fang, Distinguishing the Topological Charge of Vortex Beam via Fourier Back Plane Imaging with Chiral Windmill Structure, New J. Phys. (2024).
[48] M. Imada, A. Chutinan, S. Noda, and M. Mochizuki, Multidirectionally distributed feedback photonic crystal lasers, Phys. Rev. B **65**, 195306 (2002).
[49] S. Sun et al., Tunable plasmonic bound states in the continuum in the visible range, Phys. Rev. B **103**, 045416 (2021).